\documentclass[prd,superscriptaddress,onecolumn,showkeys]{revtex4}
\usepackage{textcomp}
\usepackage{eurosym}
\usepackage{amsfonts}
\usepackage{array}
\usepackage{amsthm}
\usepackage{bm}
\usepackage{palatino}
\usepackage[colorinlistoftodos]{todonotes}
\usepackage{mathpazo}
\usepackage{supertabular}
\usepackage{subfig}
\usepackage{amssymb}
\usepackage{eurosym}
\usepackage{amsmath}
\usepackage{epsfig}
\usepackage{graphics}
\usepackage{changes}
\usepackage{color}
\usepackage{graphicx}
\usepackage[colorlinks=true,
            linkcolor=blue,
           urlcolor=black,
           citecolor=black]{hyperref}

\def\be{\begin{equation}}
\def\ee{\end{equation}}
\def\beq{\begin{eqnarray}}
\def\eeq{\end{eqnarray}}

\def\bes{\begin{eqnarray}}
\def\ees{\end{eqnarray}}

\begin{document}
\title{Tunneling Analysis Under the Influences of Einstein-Gauss-Bonnet Black Holes Gravity Theory}

\author{Riasat Ali}
\email{riasatyasin@gmail.com}
\affiliation{Department of Mathematics, GC,
University Faisalabad Layyah Campus, Layyah-31200, Pakistan}

\author{Muhammad Asgher}
\email{m.asgher145@gmail.com}
\affiliation{Department of Mathematics, The Islamia
University of Bahawalpur, Bahawalpur-63100, Pakistan}

\begin{abstract}
We consider an equation of motion for Glashow-Weinberg-Salam model and
apply the semiclassical Hamilton-Jacobi process and WKB approximation in order to
%understand the unified theory of gauge of
%the electromagnetic and weak interaction. We
compute the tunneling
probability of W-bosons in the background of electromagnetic field to
analyze the quantum gravity effects of charged black hole(BH) in
Einstein-Gauss-Bonnet gravity theory. After this, we examine the
quantum gravity influences on the generalized Lagrangian field equation. We make clear that quantum gravity effects leave the
remnants on the tunneling radiation becomes non-thermal. Moreover, we analyze the graphical behavior of
quantum gravity influences on corrected Hawking temperature for spin-$1$ particles for charged BHs.
\end{abstract}


\date{\today}

\maketitle

\section{Introduction}

%The consider of quantum gravity mechanical effects within the basis of classical
%relativity provides many concerning physical process, which act a very significant
%component to realize the quantum gravity possibilities.
In quantum theory, the BH
radiation as a solution of Hawking \cite{R1} is one of the significant
physical phenomena. During the study of radiation phenomenon, the investigators effort to combine
the gravitation within quantum mechanics and BH thermodynamics \cite{R2, R3}.
In order to analyze the BHs radiation phenomena, many methods have been
suggested in the literature. Many researchers have observed these radiation for the
different well-known BHs \cite{R4}-\cite{R8}.
The generalized uncertainty principle (GUP) relation plays a critical role in understanding the BHs nature,
while the quantum effect may be studied as critical effect around the BH horizon.
By applying the tunneling phenomenon incorporating GUP
relation the thermodynamical  properties of BHs have been analyzed \cite{R9}.
The tunneling radiation from quartic and cubic BHs through the GUP corrected
fermion and scalar particles have been investigated \cite{R10} as well as the
author's studied that BHs temperature depend on the tunneling particles properties
and also observed the fermion and scalar particles temperature which appear alike
in both interactions.

Using Hamilton-Jacobi method, the significance of field equation has been studied by considering
electromagnetic interactions \cite{T1}-\cite{T4}. The massive bosons
tunneling in both electric and magnetic fields and Hawking temperatures have been computed. The Hawking radiation as a
semi-classical tunneling from BHs and black ring have
been investigated. In this paper, by using the WKB approximation, authors analyzed the
Lagrangian field equation for charged bosons.
The gravitational effects on BH radiation as well as the instability and stability of BH have
been analyzed and concluded that the rotation parameter effects the tunneling radiation.

In recent years, the investigators put a great interest to study the
BHs thermal properties, especially with a spacetime having charged and
cosmological constant. The main aim of this paper is to analyze the
tunneling radiation phenomena with GUP effects with charged and
coupling constant BHs from higher dimensions.
Firstly, we assume
a $D$-dimensional BH with charged and coupling constant. Then by applying
the Hamilton Jacobi technique, we compute the boson tunneling for
spin-$1$ particles. Moreover, we have observed the modified Hawking
temperature for spin-$1$ boson particles. Further, we assume the graphical analysis of
quantum gravity effects on modified Hawking temperature for spin-$1$
particles for charged BHs. The Hawking radiation process for the various
spin particles has been analyzed in literature \cite{t5}-\cite{t7}.
Furthermore, it has been analyzed that the Hawking radiation for various
spin of particles.

The singularity of spacetime is a causal contact with a outside
observer and the relevant geometry can not be explained by
semiclassical or classical theories. So, to have a logically
predictive ability, the singularity must be causally unconnected
from distance observer. This is indeed the type for the space
Schwarzschild result, where the horizon ignore the singularity
from the distance observer. Simply their
are also general relativity solutions which may comprise naked
singularity (singularity without the event horizon).
The creation of such naked singularity in the physical existence
would invalidate applicability semiclassical and
classical physics. Therefore, naked singularity is unsuitable
feature of general relativity.

The paper is sketched as follows, Sec. \textbf{II }contains the line
element information of D-dimensional BHs. Subsection \textbf{II.A}, \textbf{II.B}
and \textbf{II.C} represents the study of boson tunneling in
the BHs for $3$, $4$ and D-dimension spaces.
In Sec. \textbf{III}, we study the graphical analysis of quantum gravity effects on
Hawking temperature for spin-$1$ particles. In Sec. \textbf{IV},
we resume the conclusions of our work.

\section{Einstein-Gauss-Bonnet Black hole gravity theory}

The result of Einstein equation will not consist any naked singularity.
This result is the generalization of the Reissner-Nordstr\"{o}m
BHs in general relativity. We analyze the charged solution of D-dimension BHs in
Einstein-Gauss-Bonnet gravity theory.
We study the Lagrangian equation with Einstein-Gauss-Bonnet (EGB) gravity theory,
\begin{equation}
\L=\hat{\beta}(R^2+R_{abcd}R^{abcd}-4R_{ab}R^{ab})+R.\nonumber
\end{equation}
The BH of
spherically symmetric solution of this gravity
theory in D-dimension spacetime is  of the form \cite{D1}
\begin{equation}
ds^{2}=-G(r)dt^{2}+\frac{1}{G(r)}dr^{2}+r^2d{\Omega}^2_{d-2},\label{aa}
\end{equation}
with
\begin{equation}
G(r)=\frac{r^2}{4\beta}[1-\sqrt{1-\frac{8\beta q^2}{r^{2d-4}}+\frac{16\beta M}{r^{d-2}}}]+1, \nonumber
\end{equation}
where $q$ and $M$ are BH charge and mass, the constant $\beta$ is associated to the cupeling
constant as $(d-3)(d-4)\hat{\beta}/2$.
We represent the metric by $d{\Omega}^2_{d-2}$ for the unit $d=3$ sphere of area $A_{1}$.
The just nonzero element of the vector electromagnetic potential has the form
\begin{equation}
A_{t}(r)=-\frac{q}{r^{d-3}}.
\end{equation}
Here, $q$ and $M$ are associated to the BH Arnovwitt-Deser-Misner
charge$(\hat{q})$ and mass $\hat{M}$ as
\begin{equation}
q^2=\frac{2(d-3)}{d-2}\hat{q}^2,~~~
M=\frac{8\pi}{(d-2)A_{d-2}}\hat{M}.\nonumber
\end{equation}
\subsection{3-Dimensional Charged Black Hole}
The spherically symmetric solution of BH in $3$-dimensional spacetime is of the form
\begin{eqnarray}
ds^{2}&=&-Gdt^{2}+\frac{1}{G}dr^{2}+r^2d\theta^{2}
.\nonumber\label{3D}
\end{eqnarray}

We focus on the effect of the quantum gravity on the
boson tunneling from the BHs in EGB gravity theory.
Firstly, we apply the GUP-corrected Lagrangian field
equation for the massive charged vector field given
by \cite{D3}
\begin{eqnarray}
&&\partial_{\mu}(\sqrt{-g}\psi^{\nu\mu})+\sqrt{-g}\frac{m^2}{\hbar^2}\psi^{\nu}
+\sqrt{-g}\frac{i}{\hbar}A_{\mu}\psi^{\nu\mu}
+\sqrt{-g}\frac{i}{\hbar}eF^{\nu\mu}\psi_{\mu}
+\alpha\hbar^{2}\partial_{0}\partial_{0}\partial_{0}(\sqrt{-g}g^{00}\psi^{0\nu})\nonumber\\
&&-\alpha\hbar^{2}\partial_{i}\partial_{i}\partial_{i}(\sqrt{-g}g^{ii}\psi^{i\nu})=0,\label{A1}
\end{eqnarray}
where $\psi^{\mu\nu},~m$ and $g$ are the anti-symmetric tensor, particle mass and
coefficient matrix determinant, respectively.
The $\psi^{\mu\nu}$ can be defined as
\begin{equation}
\psi_{\nu\mu}=(1-\alpha\hbar^{2}\partial^{2}_{\nu})\partial{\nu}
\psi_{\mu}-(1-\alpha\hbar^{2}\partial^{2}_{\mu})\partial{\mu}\psi_{\nu}
+(1-\alpha\hbar^{2}\partial^{2}_{\nu})\frac{i}{h}eA_{\nu}\psi_{\mu}
-(1-\alpha\hbar^{2}\partial^{2}_{\mu})\frac{i}{h}eA_{\mu}\psi_{\nu}.\nonumber
\end{equation}
Here, $A_\mu$, $e$ and $\alpha$ are denoted as the BH potential, charged of boson
 particle and gravity parameter, respectively.
The $\psi$ components can be computed as
\begin{equation}
\psi^{0}=\frac{-1}{G}\psi_{0},~~~ \psi^{1}=G\psi_{1},~~ \psi^{2}=\frac{1}{r^2}\psi_{2},~~
\psi^{01}=-\psi_{01},~~~
\psi^{02}=-\frac{1}{Gr^2}\psi_{02},~~~
\psi^{12}=\frac{G}{r^2}\psi_{12}.\nonumber
\end{equation}
The $\psi_{\nu}$ function of the boson particle is defined as \cite{D4}
\begin{equation}
\psi_{\nu}=c_{\nu}\exp\left[\frac{\iota}{h}I_{0}(t,r,\theta)+
\sum_{i=1}^{n} h^{i}I_{i}(t,r,\theta)\right],\label{A2}
\end{equation}
where c is a constant and $I_{0}(t,r,\theta)$ is the classical action of particle.
Substituting the $\psi_{\nu}$ field function from
Eq. (\ref{A2}) into the field Eq. (\ref{A1}) for the lowest order of $h$,
we obtain the corrected Hamilton-Jacobi equations as follows;
\begin{eqnarray}
&&c_{1}(\partial_{0}I_{0})(\partial_{1}I_{0})+\alpha c_{1}
(\partial_{0}I_{0})^{3}(\partial_{1}I_{0})-c_{0}(\partial_{1}I_{0})^{2}
-\alpha c_{0}(\partial_{1}I_{0})^4+c_{1}eA_{0}(\partial_{1}I_{0})+c_{1}\alpha eA_{0}(\partial_{0}I_{0})^{2}(\partial_{1}I_{0})
+\frac{1}{Gr^2}\nonumber\\&&[c_{2}(\partial_{0}I_{0})(\partial_{2}I_{0})+\alpha c_{2}
(\partial_{0}I_{0})^3(\partial_{2}I_{0})-c_{0}(\partial_{2}I_{0})^2
-\alpha c_{0}(\partial_{2}I_{0})^4
+c_{2}eA_{0}(\partial_{2}I_{0})+
c_{2}\alpha eA_{0}(\partial_{0}I_{0})^2(\partial_{2}I_{0})]-\nonumber\\&&\frac{m^2c_{0}}{G}=0,\label{31}\\
&&c_{1}(\partial_{0}I_{0})^2+\alpha c_{1}
(\partial_{0}I_{0})^4-c_{0}(\partial_{0}I_{0})(\partial_{1}I_{0})-
\alpha c_{0}(\partial_{0}I_{0})(\partial_{1}I_{0})^{3}+c_{1}eA_{0}(\partial_{0}I_{0})
+\alpha c_{1}eA_{0}(\partial_{0}I_{0})(\partial_{1}I_{0})^2-\frac{G}{r^2}\nonumber\\&&
[c_{2}(\partial_{1}I_{0})(\partial_{2}I_{0})+\alpha c_{2}
(\partial_{1}I_{0})^3(\partial_{2}I_{0})-c_{1}(\partial_{2}I_{0})^{2}
-\alpha c_{1}(\partial_{2}I_{0})^{4}]
+eA_{0}[c_{1}(\partial_{0}I_{0})+\alpha c_{1}
(\partial_{0}I_{0})^3-c_{0}(\partial_{1}I_{0})-\nonumber\\&&\alpha c_{0}(\partial_{1}I_{0})^{3}+eA_{0}c_{1}
+\alpha c_{1}eA_{0}
(\partial_{0}I_{0})^{2}]+m^2 G c_{1}=0,\label{32}\\
&&{\frac{1}{Gr^2}}\left[c_{2}(\partial_{0}I_{0})^2+\alpha c_{2}
(\partial_{0}I_{0})^{4}-c_{0}(\partial_{0}I_{0})(\partial_{2}I_{0})
-\alpha c_{0}(\partial_{0}I_{0})(\partial_{2}I_{0})^3+c_{2}eA_{0}(\partial_{0}I_{0})\right.
+\alpha c_{2}eA_{0}(\partial_{0}I_{0})^{3}\left.\right]
+\frac{G}{r^2}\nonumber\\&&[c_{2}(\partial_{1}I_{0})^2+\alpha c_{2}
(\partial_{1}I_{0})^{4}-c_{1}(\partial_{1}I_{0})(\partial_{2}I_{0})
-\alpha c_{1}(\partial_{1}I_{0})(\partial_{2}I_{0})^3]
+\frac{m^2 c_{2}}{r^2}+\frac{eA_{0}}{Gr^2}[c_{2}(\partial_{0}I_{0})+
\alpha c_{2}(\partial_{0}I_{0})^3-c_{0}(\partial_{0}I_{0})\nonumber\\&&-\alpha c_{0}(\partial_{0}I_{0})^3+c_{2}eA_{0}+\alpha eA_{0}(\partial_{0}I_{0})^2]=0.\label{33}
\end{eqnarray}
We can take the classical action of particle in the following form
\begin{equation}
I_{0}=-(E-j\omega)t+W(r)+K\theta,
\end{equation}
where $\hat{E}=E-j\omega$, $E$, $j$ and $\omega$ shoe the energy of particle,
angular momentum and angular velocity, respectively. From Eqs. (\ref{31})-(\ref{33}),
the field equation can be expressed as
\begin{equation}
U(c_{0},c_{1},c_{2})^{T}=0,
\end{equation}
where $U$ is a $3\times3$ ordered of matrix and its elements are given below\\
\begin{eqnarray}
U_{00}&=&-[\dot{W}^2+\alpha \dot{W}^4]-\frac{1}{Gr^2}[K^2+\alpha K^4]-\frac{m^2}{G},\nonumber\\
U_{01}&=&-[\Hat{E}+\alpha\Hat{E}^3-eA_{0}-\alpha eA_{0}\Hat{E}^2]\dot{W},\nonumber\\
U_{02}&=&-\frac{1}{Gr^2}\Hat{E}+\alpha\Hat{E}^3-eA_{0}-\alpha eA_{0}\Hat{E}^2]K,\nonumber\\
U_{10}&=&-[\dot{W}+\alpha \dot{W}^3]\Hat{E}+eA_{0}[\dot{W}+\alpha\dot{W}^3],\nonumber\\
U_{11}&=&-[\Hat{E}^2+\alpha \Hat{E}^4-2eA_{0}\Hat{E}-\alpha eA_{0}\Hat{E}^3
-+(eA_{0})^2+(eA_{0})^2\dot{W}^{2}]-\frac{G}{r^2}[K^2+\alpha K^4]-m^2G,\nonumber\\
U_{12}&=&\frac{G}{r^2}[\dot{W}+\alpha \dot{W}^3]K,\nonumber\\
U_{20}&=&\frac{1}{Gr^2}[K+\alpha K^3]\Hat{E}+\frac{eA_{0}}{Gr^2}[K+\alpha K^3],\nonumber\\
U_{21}&=&-\frac{G}{r^2}[K+\alpha K^3]\dot{W},\nonumber\\
U_{22}&=&-\frac{1}{Gr^2}[\Hat{E}^2+\alpha\Hat{E}^4-2eA_{0}\Hat{E}-2\alpha eA_{0}\Hat{E}^3
+(eA_{0})^2+(eA_{0})^2\alpha\Hat{E}^2]+\frac{G}{r^2}[\dot{W}^2+\alpha \dot{W}^4]-\frac{m^2}{r^2},\nonumber
\end{eqnarray}
where $\dot{W}=\partial_{r}I_{0}$ and
$K=\partial_{\theta}I_{0}$.
For the result of non-trivial solution, we take $U=0$ and get
\begin{equation}
ImW^{\pm}=\pm \int\frac{\sqrt{(\Hat{E}-eA_{0})^{2}+X_{1}[1+\alpha\frac{X_{2}}{X_{1}}]}}{G}dr,
=\pm i\pi\frac{(\Hat{E}-eA_{\psi})}{2\kappa(r_{+})}[1+\alpha\Xi],\label{RA}
\end{equation}
where $W^{-}$ and $W^{+}$ represent to the incoming and outgoing trajectories of boson particle,
respectively, and the values of $X_{1}$ and $X_{1}$ are given as
\begin{equation}
X_{1}=-2eA_{0}\Hat{E}+Gm^2,~~~
X_{2}=\Hat{E}^4-2eA_{0}\Hat{E}^3+(eA_{0})^2\Hat{E}^2-G\dot{W}^{4}.\nonumber
\end{equation}
where ${\kappa}(r_{+})$ is the standard BH surface gravity and its explicit
form is
\begin{equation}
{\kappa}(r_{+})=\frac{4\beta q^2-8\beta M+r\sqrt{8\beta (2M-q^2)}-r^2}{4\beta \sqrt{r^2+16\beta M-8\beta q^2}}.
\end{equation}
Hence, the boson particle tunneling can be obtained as
\begin{eqnarray}
\Gamma=\exp\left[-2\pi\frac{4\beta \sqrt{r^2+16\beta M-8\beta q^2}(\Hat{E}-eA_{0})(1+\alpha \Xi)}
{4\beta q^2-8\beta M+r\sqrt{8\beta (2M-q^2)}-r^2}\right].
\end{eqnarray}
The gravitational Hawking temperature of the boson particle for the BH can be written as
\begin{equation}
{T}_{H}=\frac{4\beta q^2-8\beta M+r_{+}\sqrt{8\beta (2M-q^2)}-r_{+}^2}{8\pi
\beta \sqrt{r_{+}^2+16\beta M-8\beta q^2}}[1+\alpha \Xi].\label{TH3}
\end{equation}
The boson particle tunneling depends on BH potential,
BH charged, cupeling constant, BH radius, BH mass, particle velocity,
particles energy and particle momentum as well as quantum gravity.
While, the temperature depends on properties of BH and quantum
gravity parameter. This result shows that the corrected
temperature of the boson charged particle is higher than the
absolute temperature. Moreover, it indicates that the
modified temperature in Eq. ({\ref{TH3}}) depends upon not only the
BH properties but also the quantum gravity parameter.
\subsection{4-Dimensional Charged Black Hole}
The spherically symmetric solution of BH in $4$-dimensional spacetime is of the form
\begin{eqnarray}
ds^{2}&=&-Gdt^{2}+\frac{1}{G}dr^{2}+r^2d\theta^{2}+r^2sin^2\theta d\psi.\nonumber\label{aa}
\end{eqnarray}
For the requiring component of the electromagnetic vector potential($A_{0}$) and quantum gravity,
the explicit form of the tunneling probability can be computed by the above way:
\begin{eqnarray}
\Gamma=\exp\left[-2\pi\frac{4\beta \sqrt{r^4-8\beta Q^2+16M\beta r}(\Hat{E}-eA_{0})(1+\alpha \Xi)}
{r^3\sqrt{r^4+16\beta Mr-8\beta Q^2}-4\beta M-r^3}\right].
\end{eqnarray}
Furthermore, observing the quantum gravity terms, we calculate the
Hawking temperature of the $4$-dimensional charged BH in EGB gravity theory as follows
\begin{equation}
{T}_{H}=\frac{r_{+}^3\sqrt{r_{+}^4+16\beta Mr_{+}-8\beta Q^2}-4\beta M-r_{+}^3}
{8\pi \beta \sqrt{r_{+}^4-8\beta Q^2+16M\beta r_{+}}}[1+\alpha \Xi].\label{TH4}
\end{equation}
This solution indicates that the temperature with gravity of the charged
boson particle is higher than the temperature without gravity and depends
on the BH properties. Also, it is similar from that of the fermion and scalar particles.
\subsection{D-Dimensional Charged Black Hole}
The spherically symmetric solution of BH in $D$-dimension spacetime is of the form
\begin{equation}
ds^{2}=-G(r)dt^{2}+\frac{1}{G(r)}dr^{2}+r^2d{\Omega}^2_{d-2}.
\end{equation}
The tunneling rate of the $D$-dimension charged BH in EGB gravity theory can be calculated as follows
\begin{eqnarray}
\Gamma&=&\exp[ (-2\pi\frac{2\beta M(D-1)r^{2-D}+(4-2D)Q^2r^{5-2D}}
{2\sqrt{1-8\beta Q^2r^{4-2D}+16\beta Mr^{1-D}}}
+\frac{r(1-\sqrt{1-8\beta Q^2r^{4-2D}+16\beta Mr^{1-D}})}{4\beta})\nonumber\\
&&(\Hat{E}-eA_{0})(1+\alpha\Xi)].
\end{eqnarray}
The Hawking temperature of the D-dimension charged BH in EGB gravity theory can be calculated as
\begin{equation}
{T}_{H}=\left[\frac{2\beta M(D-1)r_{+}^{2-D}+(4-2D)Q^2r_{+}^{5-2D}}{4\pi \sqrt{1-8\beta Q^2r_{+}^{4-2D}+16\beta Mr_{+}^{1-D}}}
+\frac{r_{+}(1-\sqrt{1-8\beta Q^2r_{+}^{4-2D}+16\beta Mr_{+}^{1-D}})}{8\pi \beta }\right][1+\alpha \Xi].\label{THD}
\end{equation}
This solution indicates that the boson particle temperature depends
on the BH dimension and properties. Also, it is different from the three
and four dimensional BHs temperature of the boson particle.
\section{Graphical Stability Analysis}

The stability of the different-dimensional charged BHs in EGB gravity theory can be
studied by the tunneling radiation. If the Hawking temperature is positive, then
the BHs is stable or else it is unstable.
For this purpose, to study the stability of the different-dimensional charged BHs in
EGB gravity theory, we firstly compute its modified temperature.
The temperature graphical analysis at constant charged$(Q)$, constant arbitrary
parameter$(\Xi)$, mass BH constant, coupling constant BH radius constant and vary
the quantum gravity can be obtained by applying the following relations
({\ref{TH3}}, {\ref{TH4}}, {\ref{THD}}).

From Fig. 1, we see the approximation region $0<r_{+}<5$, it is stable.
Since, we can say that the $3$-dimensional charged BHs in EGB gravity theory become
unstable to the presence of the GUP effect. However, if $\alpha=100-300$ then
corrected temperature is positive. Hence, the 3-dimensions charged BHs in
EGB gravity theory undergoes a initial to stabile. We can say that the BH
will be undergo to unstable in the presence of large the quantum gravity and also temperature
will be decrease.

In Fig. 2, our results show that the temperature of the
3-dimensions charged BHs in EGB gravity theory depend on the BH properties as well
as BH mass and boson tunneling particles in the presence of the quantum gravity.
It is worth to observe that the boson particle temperatures calculated through
boson tunneling phenomenon. The modified temperature are lower than the BH stable
and the corrected temperature are higher than the BH unstable. The corrected
temperature are lower than the small quantum gravity and the corrected temperature
are higher than the large quantum gravity.

In Fig. 3, the absence of the GUP parameter, i.e. $\alpha=0$, the corrected
temperature are obtained to the standard temperature.
in the presence of the quantum gravity parameter, if the quantum gravity of
the D-dimensions charged BHs in EGB gravity theory is positive, the BH becomes stable
in the approximation range $0<r_{+}<5$  but, in the quantum gravity absence, the
stability property of the BH does not change and the BH unstable in the approximation range
$5<r_{+}<\infty$. When we ignore quantum gravity parameter then BH is more stable.
Our results shows that quantum gravity effects leave the
remnants on the boson tunneling radiation implies non-thermal.

Finally, we observe that the quantum gravity effect on $3$, $4$ and $D$ dimensions
charged BHs in EGB gravity theory of temperature decreases and increases with increasing
horizon $r_{+}$ and also observe the physical significance of
this temperature to see the under the more and lower influence of quantum gravity
on $T'_{H}$ by observing the BH totally instable and may be stable(i.e., quantum gravity inversely proportional to stability).\\\\\\

\begin{figure}\begin{center}
\epsfig{file=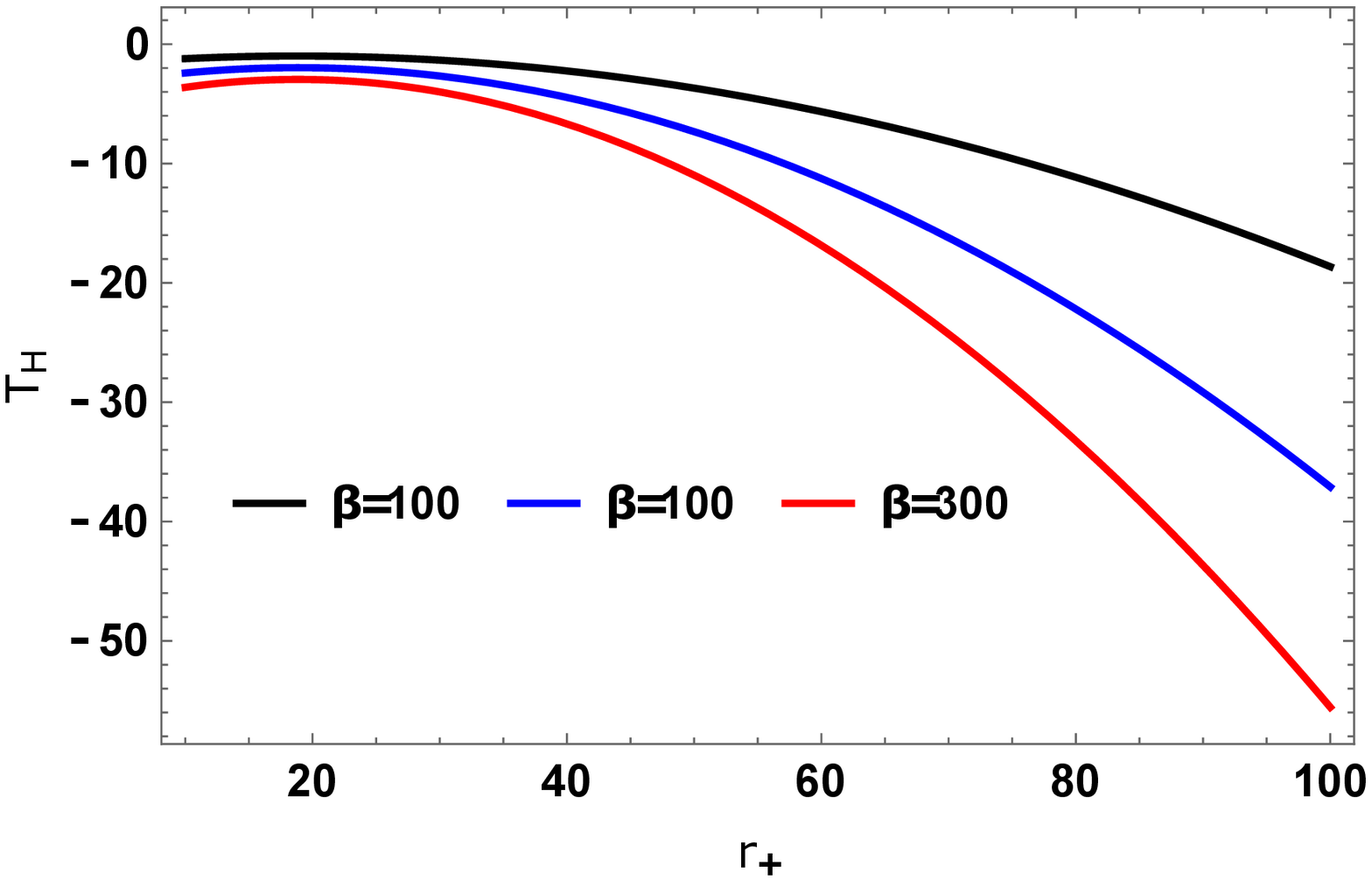, width=0.5\linewidth}
\caption{$T_{H}$ versus $r_{+}$ for $M =100,~q=5,~~\beta = 1,$ and $\Xi=1.$}
\end{center}
\begin{center}
\epsfig{file=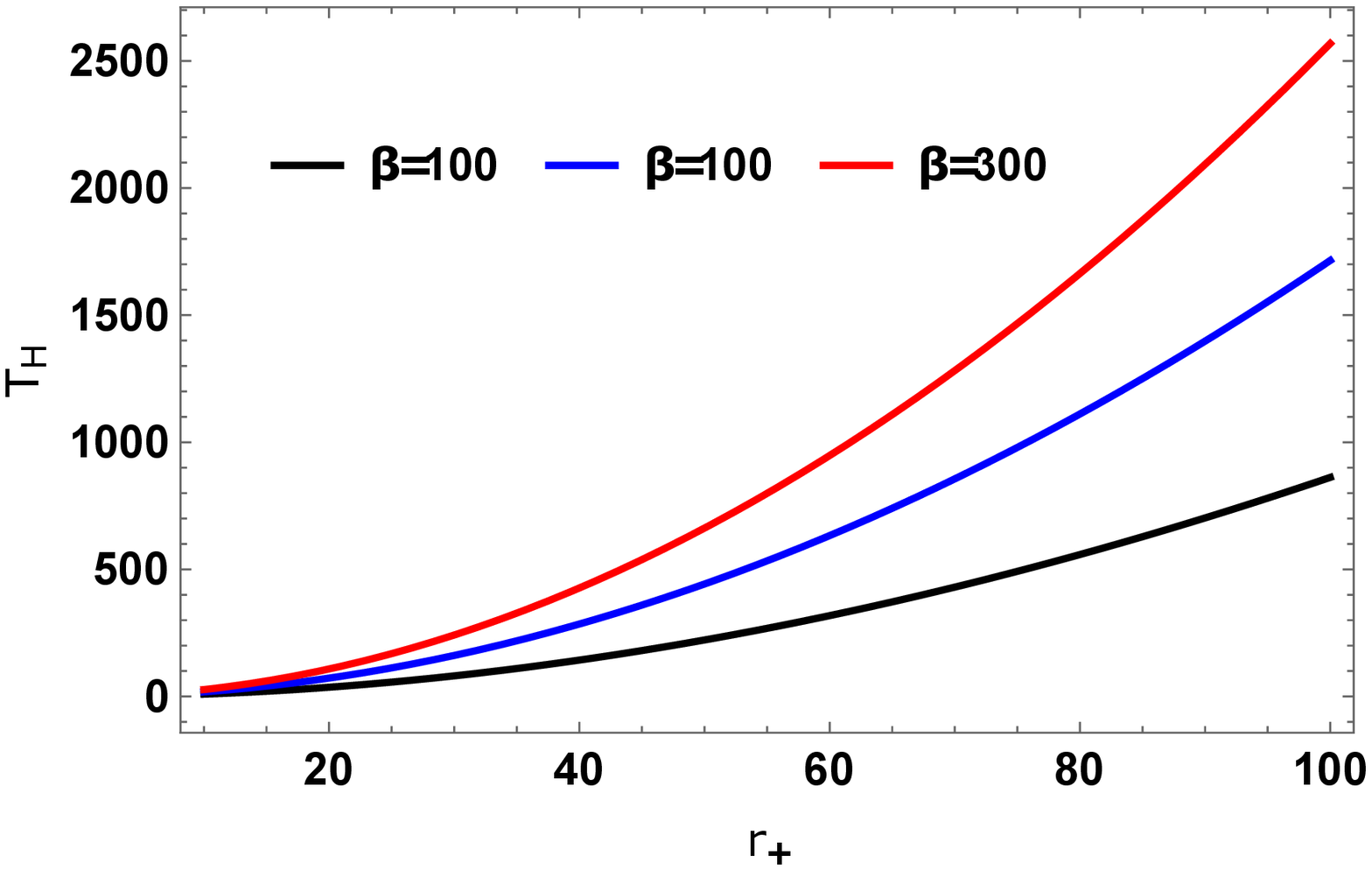, width=0.5\linewidth}
\caption{$T_{H}$ versus $r_{+}$ for $M =100,~q=5,~~\beta = 1,$ and $\Xi=1$}
%\end{center}\end{figure}
%\begin{figure}\begin{center}
\epsfig{file=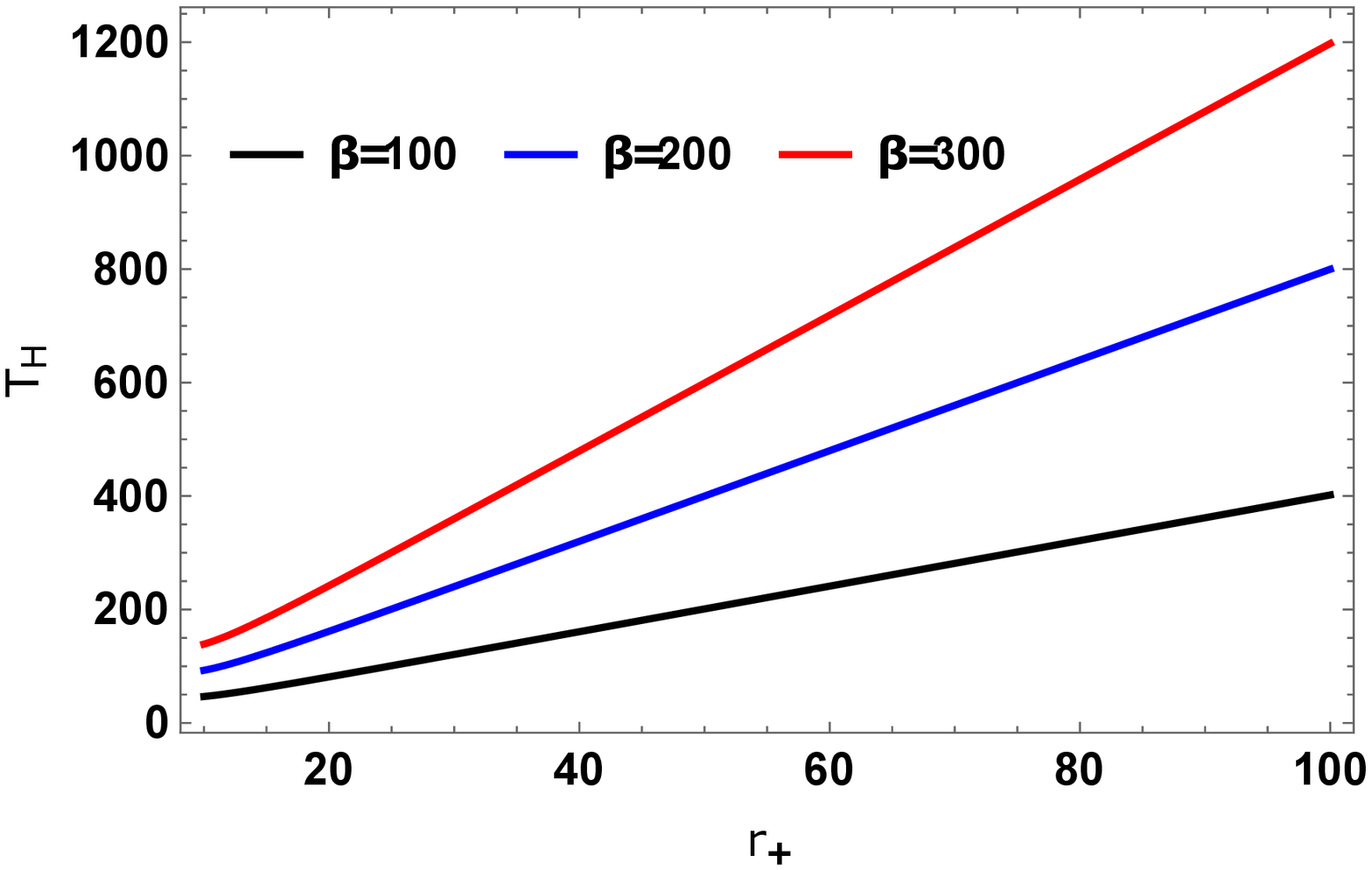, width=0.5\linewidth}
\caption{$T_{H}$ versus $r_{+}$ for $M =100,~q=5,~d=5,~\beta = 1,$ and $\Xi=1$.}
\end{center}
\end{figure}

\section{Conclusions}

In this paper, we have analyzed the quantum gravity effect on the temperature of the
$3$, $4$ and $D$ dimensional charged BHs in EGB gravity theory in the context of
the boson tunneling phenomenon of the massive charged spin-$1$ particle.
Our results show that the modified temperature of the 3, 4 and  D dimensional
charged BHs in EGB gravity theory depend on the BH geometry, but also the quantum
tunneling gravity parameter. It is deserving to mention that the modified temperatures
computed through boson particle tunneling are completely similar from each other
particles such as scalar and fermion.
The corrected temperatures are higher than the absolute temperatures in all different
3, 4 and  D-dimensions charged BHs. In the quantum gravity effect absence, i.e. $\alpha=0$, the corrected temperatures are
obtained to the absolute temperatures in all cases.

In this study, we observe the stability property of the BHs by using the corrected
temperature of the charged massive boson particle. The $3$-dimensional charged BH undergoes
an initial to stabile and BH will be undergo to unstable when large quantum gravity and
also temperature will be decrease(Fig. 1). The temperature are lower than the $4$ and $D$ dimensional
BH stable and the temperature are higher than the BH unstable. The temperatures are
lower than the small quantum gravity and the temperatures are higher than the large
quantum gravity (figures 2 and 3). In the quantum gravity absence then the BHs are more stable.
The boson particle temperature depends on the BH dimensions and properties.
Also, it is different from the three dimension or higher dimensions BHs temperature
of the boson particle.

Finally, we can analyze that the quantum gravity of the
boson tunneling particles play a significant role in understanding the BHs evolution
such that it can study radiation on the BHs final stage. From graphs, we see that the quantum gravity effect on 3, 4 and D dimensions
charged BHs in EGB gravity theory of temperature decreases and increases with increasing
horizon $r_{+}$.


\begin{thebibliography}{90}

\bibitem{R1} Hawking SW, (1975). Particle creation by black holes, Commun. Math. Phys. {\bf 43}(1975), p. 199-220.
https://doi.org/10.1007/BF02345020.

\bibitem{R2} Hawking SW, (1974). Black hole explosions?, Nature {\bf 248}(1974), p. 30-31.
https://doi.org/10.1038/248030a0.

\bibitem{R3} Hawking SW, (1976). Black holes and thermodynamics, Phys. Rev. D {\bf 13}(1976), p. 191.
https://doi.org/10.1103/PhysRevD.13.191.

\bibitem{R4}  Sharif M, Javed W, (2012). Fermions Tunneling from Charged Accelerating
and Rotating Black Holes with NUT Parameter, Eur. Phys. J. C {\bf 72}(2012), p. 1-10.
https://doi.org/10.1140/epjc/s10052-012-1997-y.

\bibitem{R5}  Sharif M, Javed W, (2013). Thermodynamics of a Bardeen black hole in
noncommutative space, Can. J. Phys. {\bf 91}(2013), p. 43.
https://doi.org/10.1139/p11-089.

\bibitem{R6}  Javed W, Abbas G,  Ali R, (2017). Charged vector particle tunneling from a pair of accelerating
and rotating and 5D gauged super-gravity black holes, Eur. Phys. J. {\bf C 77}(2017), p. 296.
https://doi.org/10.1140/epjc/s10052-017-4859-9.

\bibitem{D3}  Javed W, Ali R, Babar R, \"{O}vg\"{u}n, A, (2020). Tunneling of massive vector particles under influence of
quantum gravity, Chinese Physics {\bf C 144}(2020), p. 015104.

\bibitem{12} Javed W, Babar R, (2019). Vector particles tunneling in the background of quintessential field
involving quantum effects, Chinese Journal of Physics, \textbf{61}(2019), p. 138-254.
https://doi.org/10.1016/j.cjph.2019.08.016

\bibitem{D4} Javed W,  Ali R, Babar R, \"{O}vg\"{u}n, A, (2019). Tunneling of massive vector particles from types of BTZ-like
black holes, Eur. Phys. J. Plus {\bf 134}(2019), p. 511.
https://doi.org/10.1140/epjp/i2019-12877-5.

\bibitem{R7}  Javed W, Ali R, Abbas G, (2018). Charged vector particles tunneling from black ring
and 5D black hole, Can. J. Phys. {\bf 97}(2018), p. 176.
https://doi.org/10.1139/cjp-2018-0270.

\bibitem{R8}  \"{O}vg\"{u}n A, Javed W, Ali R, (2018). Tunneling Glashow-Weinberg-Salam Model Particles from
Black Hole Solutions in Rastall Theory,  Advance in High Energy Physics. {\bf 2018}(2018), p. 11.
https://doi.org/10.1155/2018/3131620.

\bibitem{R9}  Kanzi S, Sakalli I, (2019). GUP modified Hawking radiation in bumblebee gravity, Nucl. Phys. {\bf B 946}(2019), p. 114703.
https://doi.org/10.1016/j.nuclphysb.2019.114703.

\bibitem{R10}  Javed W, Babar R, \"{O}vg\"{u}n A, (2019). Hawking radiation from cubic and quartic black holes via tunneling of GUP corrected scalar and fermion particles, Mod. Phys. Lett. {\bf A 34}(2019), p. 1950057;
https://doi.org/10.1142/S0217732319500573


\bibitem{T1}  Ali R, Bamba K, Shah SAA, (2019).  Effect of Quantum Gravity on the Stability of
Black Holes, Symmetry {\bf 631}(2019), p. 11.
https://doi.org/10.3390/sym11050631.

\bibitem{T2}  Ali R, et al (2020). Stability Analysis of Charged Rotating Black Ring, Symmetry. {\bf 1165}(2020), p. 12.
 https://doi.org/10.3390/sym12071165.

\bibitem{T3}  Ali R, Asgher M,  Malik MF, (2020). Gravitational analysis of neutral regular black hole
in Rastall gravity, Mod. Phys. Lett. {\bf A 35}(2020), p. 2050225.
https://doi.org/10.1142/S0217732320502259.

\bibitem{T4}  Ali R, et al (2021). Tunneling under the influence of quantum
gravity in black rings, Int. J. Mod. Phys. {\bf D 30}(2021), p. 2150002.
https://doi.org/10.1142/S0218271821500024.

\bibitem{t5}  Li XQ, Chen GR, (2015). Massive vector particles tunneling from Kerr and Kerr-Newman black holes, Phys. Lett. {\bf B 751}(2015), p. 34-38.
https://doi.org/10.1016/j.physletb.2015.10.007.

\bibitem{t6}  Babar R, Javed W, \"{O}vg\"{u}n A, (2020). Effect of the GUP on the Hawking radiation of black hole in 2+1 dimensions with quintessence and charged BTZ-like magnetic black hole, Mod. Phys. Lett. {\bf A 35}(2020), p. 2050104.
https://doi.org/10.1142/S0217732320501047.

\bibitem{t7} Yale A, (2011). Exact Hawking radiation of scalars, fermions, and bosons using the tunneling
method without back-reaction, Phys. Lett. {\bf B 697}(2011), p. 398.
https://doi.org/10.1016/j.physletb.2011.02.023.

\bibitem{D1}  Ghosh R, Fairoos C, Sarkar, S, (2019). Overcharging higher curvature black holes,  Phys. Rev. D {\bf 100}, p. 124019.
https://doi.org/10.1103/PhysRevD.100.124019.
\end{thebibliography}
\end{document}